\documentstyle[preprint,aps,epsfig]{revtex}
\tightenlines
\begin{document}

\preprint{
\vbox{
\hbox{ADP-00-13/T397}
\hbox{JLAB-THY-00-08}
}}

\title{Local Duality Predictions for $x \sim 1$ Structure Functions}
\author{W. Melnitchouk}
\address{Jefferson Lab,
	12000 Jefferson Avenue,
	Newport News, VA 23606, and				\\
	Special Research Centre for the Subatomic Structure of Matter,\\
	Adelaide University, Adelaide 5005, Australia}
\maketitle

\begin{abstract}
Recent data on the proton $F_2$ structure function in the resonance
region suggest that local quark-hadron duality works remarkably well
for each of the low-lying resonances, including the elastic, to rather
low values of $Q^2$.
We derive model-independent relations between structure functions at
$x \sim 1$ and elastic electromagnetic form factors, and predict the
$x \to 1$ behavior of nucleon polarization asymmetries and the neutron
to proton structure function ratios from available data on nucleon
electric and magnetic form factors.
\end{abstract}

\newpage
The nucleon's deep-inelastic structure functions and elastic form factors
parameterize fundamental information about its quark substructure.
Both reflect dynamics of the internal quark wave functions describing
the same physical ground state, albeit in different kinematic regions.
Recent work on generalized parton distributions \cite{NFPD} has provided
a unifying framework within which both form factors and structure
functions can be simultaneously embedded.

Exploration of the structure function---form factor interface is
actually as old as the first deep-inelastic scattering experiments
themselves.
In the early 1970s the inclusive--exclusive connection was studied in
the context of deep-inelastic scattering in the resonance region and
the onset of scaling behavior.
In their pioneering work, Bloom and Gilman \cite{BG} observed that
the inclusive $F_2$ structure function at low $W$ generally follows
a global scaling curve which describes high $W$ data, to which the
resonance structure function averages.
Furthermore, the equivalence of the averaged resonance and scaling
structure functions appears to hold for each resonance, over restricted
regions in $W$, so that the resonance---scaling duality also exists
locally.

More recently, high precision data on the $F_2$ structure function from
Jefferson Lab \cite{JLAB} have confirmed the original observations of
Bloom and Gilman, demonstrating that local duality works remarkably well
for each of the low-lying resonances, including surprisingly the elastic,
to rather low values of $Q^2$.
In the context of the operator product expansion in QCD, the existence of
Bloom-Gilman duality can be attributed to the small size of higher twist
($1/Q^2$ suppressed) contributions to the structure function.
In this Letter we examine local duality for the elastic case more closely,
and derive {\em model-independent} relations between structure functions
at $x \sim 1$ and elastic electromagnetic form factors.
Using the most recent data on the nucleon electric and magnetic form
factors, we apply local duality to make quantitative predictions for the
asymptotic behavior of unpolarized and polarized structure function
ratios.

To illustrate the interplay between resonances and scaling functions,
one can observe \cite{BG,CM,JIMV} that (in the narrow resonance
approximation) if the contribution of a resonance of mass $M_R$ to the
$F_2$ structure function at large $Q^2$ is given by
$F_2^{(R)} = 2 M \nu \left( G_R(Q^2) \right)^2\ \delta(W^2~-~M_R^2)$,
then a form factor behavior $G_R(Q^2) \sim (1/Q^2)^n$ translates into
a scaling function
$F_2^{(R)} \sim (1-x_R)^{2n-1}$, where $x_R = Q^2/(M_R^2 - M^2 + Q^2)$.
On purely kinematical grounds, therefore, the resonance peak at $x_R$
does not disappear with increasing $Q^2$, but rather moves towards
$x=1$.

For elastic scattering, the connection between the $1/Q^2$ power of the
elastic form factors at large $Q^2$ and the $x \to 1$ behavior of
structure functions was first established by Drell and Yan \cite{DY}
and West \cite{WEST}.
More recently, interest in large-$x$ structure functions has arisen
in connection with the polarization asymmetry $A_1 = g_1/F_1$, and the
$F_2^n/F_2^p$ ratio, whose $x \to 1$ limits reflect mechanisms for the
breaking of spin-flavor SU(6) symmetry in the nucleon \cite{HIX}.

If the inclusive--exclusive connection via local duality is taken
seriously, one can use measured structure functions in the resonance
region at large $x$ to directly extract elastic form factors
\cite{RUJ}.
Conversely, empirical electromagnetic form factors at large $Q^2$ can
be used to predict the $x \to 1$ behavior of deep-inelastic structure
functions \cite{BG}.
To quantify this connection, we begin by noting that the elastic 
contributions to the inclusive spin-averaged structure functions can be
expressed through electric and magnetic form factors as \cite{CM}:
\begin{mathletters}
\label{SFel}
\begin{eqnarray}
F_1^{\rm el}
&=& M \tau\
    G_M^2\ \delta\left( \nu - {Q^2 \over 2M} \right),	\\
F_2^{\rm el}
&=& { 2 M \tau \over 1 + \tau } 
    \left( G_E^2 + \tau G_M^2 \right)\
    \delta\left( \nu - {Q^2 \over 2M} \right),
\end{eqnarray}
where $\tau = Q^2/4M^2$, while for spin-dependent structure functions
\cite{CM,JG1}:
\begin{eqnarray}
g_1^{\rm el}
&=& { M \tau \over 1 + \tau }
    G_M \left( G_E + \tau G_M \right)
    \delta\left( \nu - {Q^2 \over 2M} \right),		\\
g_2^{\rm el}
&=& { M \tau^2 \over 1 + \tau }
    G_M \left( G_E - G_M \right)
    \delta\left( \nu - {Q^2 \over 2M} \right)\ .
\end{eqnarray}
\end{mathletters}%
Following de R\'ujula et al. \cite{RUJ}, one can integrate Eq.(\ref{SFel})
over the targer-mass corrected scaling variable
$\xi = 2x / (1 + \sqrt{1 + x^2/\tau})$ between the pion threshold and 1,
allowing the ``localized'' moments of scaling functions to be expressed in
terms of elastic form factors.
The assumption of local elastic duality is that the area under the elastic
peak (given by integrating the right hand side of Eq.(\ref{SFel}) at a
given $Q^2$) is the same as the area under the scaling function
(at much larger $Q^2$) when integrated from the pion threshold to the
elastic point \cite{BG}.
Using the local duality hypothesis, de R\'ujula et al. \cite{RUJ}, and
more recently Niculescu et al. \cite{JLAB}, extracted the proton's $G_M$
form factor (assuming that the ratio $G_E/G_M$ is sufficiently
constrained) from resonance data on the $F_2$ structure function at
large $\xi$, finding better than $\sim 30\%$ agreement over a large range
of $Q^2$.

Applying the argument in reverse, one can formally differentiate the local
elastic duality relation \cite{BG} with respect to $Q^2$ to express the
scaling functions, evaluated at threshold, in terms of $Q^2$
derivatives of elastic form factors:
\begin{mathletters}
\label{SFdual}
\begin{eqnarray}
F_1(x=x_{\rm th}) &=& \beta\
{ dG_M^2 \over dQ^2 }\ ,					\\
F_2(x=x_{\rm th}) &=& \beta
\left\{ { G_M^2 - G_E^2 \over 2 M^2 (1+\tau)^2 }
+ { 2 \over 1 + \tau }
  \left( { dG_E^2 \over dQ^2 } + \tau { dG_M^2 \over dQ^2 }
  \right)
\right\}							\nonumber\\
&\to& 2 \beta\ { dG_M^2 \over dQ^2 }\ \ \ {\rm as}\ \tau\to\infty\ ,	\\
\label{SFdualg1}
g_1(x=x_{\rm th}) &=& \beta\
\left\{ { G_M \left( G_M - G_E \right) \over 4 M^2 (1+\tau)^2 }
+ { 1 \over 1 + \tau }
  \left( { d(G_E G_M) \over dQ^2 } + \tau { dG_M^2 \over dQ^2 }
  \right)
\right\}							\nonumber\\
&\to& \beta\ { dG_M^2 \over dQ^2 }\ \ \ {\rm as}\ \tau\to\infty\ ,	\\
\label{SFdualg2}
g_2(x=x_{\rm th}) &=& \beta\
\left\{
{ G_M \left( G_E - G_M \right) \over 4 M^2 (1+\tau)^2 }
+ { \tau \over 1 + \tau }
  \left( { d(G_E G_M) \over dQ^2 } - { dG_M^2 \over dQ^2 }
  \right)
\right\}							\nonumber\\
&\to& \beta\ { d \over dQ^2 }(G_E G_M - G_M^2)\ \ \
	{\rm as}\ \tau\to\infty\ ,
\end{eqnarray}
\end{mathletters}%
where $x_{\rm th} = Q^2 / (W^2_{\rm th} - M^2 + Q^2)$, with
$W_{\rm th} = M + m_\pi$, corresponds to the pion production threshold,
and the kinematic factor
$\beta = (Q^4/M^2) (\xi_0^2/\xi^3) (2-\xi/x)/(2\xi_0-4)$.
It is interesting to observe that asymptotically in the $Q^2 \to \infty$
limit each of the structure functions $F_1$, $F_2$ and $g_1$ is
determined by the slope of the square of the magnetic form factor,
while $g_2$ (which in deep-inelastic scattering is associated with
higher twists) is determined by a combination of $G_E$ and $G_M$.

The interpretation of the relations in Eq.(\ref{SFdual}) follows that
given by Bloom \& Gilman in the context of finite-energy sum rules
\cite{BG}.
Formulated originally by Dolen, Horn and Schmid \cite{FESR} for hadron
scattering, finite-energy sum rules relate resonance structure functions
at finite $Q^2$, averaged over appropriate intervals in $W$ (or $\nu$),
to smooth scaling functions, such as those measured in the deep-inelastic
region at much larger $Q^2$, which (modulo perturbative $\log Q^2$
corrections) depend on $x$ only.
For local elastic duality, the relevant interval over which the structure
functions are averaged is between the pion production threshold at
$x=x_{\rm th}$ and the elastic point, $x = 1$.
Clearly, in the sub-threshold region the only contribution is from elastic
scattering, which is given entirely by the elastic form factors on the
right hand side of Eq.(\ref{SFel}).

Differentiating the finite-energy sum rule relations for the elastic case
\cite{BG}, local duality then allows one to equate the right hand side of
Eq.(\ref{SFdual}), which represents the elastic contribution to the
structure functions at finite $Q^2$, with the left hand side, which
corresponds to structure functions in the scaling region.
Aside from perturbative QCD corrections, in the scaling region the latter
are functions only of $x$.

The scaling functions on the left hand side of Eq.(\ref{SFdual}) are
evaluated at $x=x_{\rm th}$, with $x_{\rm th}$ corresponding to the
particular value of $Q^2$ on the right hand side of (\ref{SFdual})
\cite{BG}.
However, since the results in the scaling limit are $Q^2$ independent,
the predictions are also valid for $x > x_{\rm th}$.
Note that in the limit $Q^2 \to \infty$ the location of the pion
threshold $x_{\rm th} \to 1$, and the kinematic factor
$\beta \to -Q^4/(2\xi_0 M^2)$.
In this limit one can explicitly verify that the right hand side of
(\ref{SFdual}) gives the correct asymptotic behavior of the structure
functions as $x \to 1$.
If $G_M(Q^2) \sim (1/Q^2)^n$ at large $Q^2$, then the right hand sides
of (\ref{SFdual}) must scale like $(1/Q^2)^{2n-1}$.
At fixed $W$, since $(1-x)$ behaves like $1/Q^2$, the $x$ dependence of
the scaling functions at large $x$ is $(1-x)^{2n-1}$, as required by
the asymptotic scaling laws \cite{DY,WEST}.

Equation (\ref{SFdual}) allows the large-$x$ behavior of structure
functions to be predicted from empirical electromagnetic form factors.
Of particular interest is the $x \to 1$ behavior of the polarization
asymmetry, $A_1$, which at large $Q^2$ is given by the ratio of
spin-dependent to spin-averaged structure functions, $A_1 = g_1/F_1$.
{}From spin-flavor SU(6) symmetry one expects, at leading twist,
$A_1 = 5/9$ for the proton, and $A_1 = 0$ for the neutron.
A number of models which incorporate SU(6) breaking, through either
perturbative or non-perturbative mechanisms \cite{HIX}, suggest
that $A_1 \to 1$ as $x \to 1$, in dramatic contrast to the SU(6)
predictions, especially for the neutron.

Using the parameterization of global form factor data from Ref.\cite{MDM},
the proton and neutron asymmetries arising from the local quark--hadron
duality relations (\ref{SFdual}) are shown in Fig.~1 as a function of $x$,
with $x$ corresponding to $x_{\rm th}$.
One sees that while for $x \alt 0.9$ (which for the pion threshold
corresponds to $Q^2 \approx 2.5$~GeV$^2$) the asymmetries are
qualitatively consistent with the SU(6) expectations, the trend
as $x \to 1$ is for both asymmetries to approach unity.
Since $x_{\rm th} \to 1$ as $Q^2 \to \infty$, this is consistent
with the operator product expansion interpretation of de R\'ujula
et al.~\cite{RUJ} in which duality should be a better approximation
with increasing $Q^2$.

Although the curves in Fig.~1 are shown for $x > 0.6$, one should note
that the region below $x = x_{\rm th} \approx 0.8$ corresponds to
$Q^2 \alt 1$~GeV$^2$, where duality is not expected to be as good an
approximation, so the reliability of the local duality predictions
there would be more questionable.
Unfortunately the current data on $A_1$ extend only out to an average
$\langle x \rangle \sim 0.5$, and are inconclusive about the $x \to 1$
behavior.
While the proton $A_1$ data do indicate a steep rise at large $x$,
the neutron asymmetry is, within errors, consistent with zero over
the measured range \cite{A1}.
It will be of great interest in future to observe whether, and at which
$x$ and $Q^2$, the $A_1$ asymmetries start to approach unity.

Another quantity of current interest is the ratio of the neutron to
proton $F_2$ structure functions at large $x$ \cite{HIX}.
There are a number of leading twist predictions for this ratio, ranging
from 2/3 in the SU(6) symmetric quark model, to 1/4 in broken SU(6)
through $d$ quark suppression \cite{SCALAR}, to 3/7 in broken SU(6) via
helicity flip suppression \cite{FJ}.
Although it is well established that the large-$x$ $F_2^n/F_2^p$ data
deviate from the SU(6) prediction, they are at present inconclusive about
the precise $x \to 1$ limit because of large nuclear corrections in the
extraction of $F_2^n$ from deuterium cross sections beyond $x \sim 0.6$
\cite{MST}.

The ratios of the neutron to proton $F_1$, $F_2$ and $g_1$ structure
functions are shown in Fig.~2 as a function of $x$, with $x$ again
evaluated at $x_{\rm th}$.
While the $F_2$ ratio varies somewhat with $x$ at lower $x$, beyond
$x \sim 0.85$ it remains almost $x$ independent, approaching the
asymptotic value $(dG_M^{n 2}/dQ^2)/(dG_M^{p 2}/dQ^2)$.
Because the $F_1$\ $n/p$ ratio depends only on $G_M$, it remains flat
over nearly the entire range of $x$ (and $Q^2$).
At asymptotic $Q^2$ the model predictions for $F_1(x\to 1)$ coincide
with those for $F_2$; at finite $Q^2$ the difference between $F_1$ and
$F_2$ can be used to predict the $x \to 1$ behavior of the longitudinal
structure function, or the $R = \sigma_L / \sigma_T$ ratio.

The spin-dependence of the proton vs. neutron duality predictions is
also rather interesting.
Since $A_1^n$ is zero for all $x$ according to SU(6), the ratio of the
neutron to proton $g_1$ structure functions is also zero in the
spin-flavor symmetric limit.
The pattern of SU(6) breaking for $g_1^n/g_1^p$ essentially follows that
for $F_2^n/F_2^p$, namely 1/4 in the $d$ quark suppression \cite{SCALAR}
and 3/7 in the helicity flip suppression \cite{FJ} scenarios.
According to local duality, the $g_1$ structure function ratio in Fig.~2
approaches the asymptotic limit in Eq.(\ref{SFdualg1}), albeit somewhat
slowly, reflecting the relatively slow approach towards unity of the
polarization asymmetry in Fig.~1.
This may indicate a larger role played by higher twists in $g_1$
compared with $F_2$, a result consistent with analyses of higher twist
corrections to moments of the $g_1$ \cite{JG1} and $F_2$ structure
functions \cite{JF2}.

It appears to be an interesting coincidence that the helicity retention
model \cite{FJ} prediction of 3/7 is very close to the empirical ratio of
the squares of the neutron and proton magnetic form factors,
$\mu_n^2/\mu_p^2 \approx 4/9$.
Indeed, if one approximates the $Q^2$ dependence of the proton and neutron
form factors by dipoles, and takes $G_E^n \approx 0$, then the structure
function ratios are all given by simple analytic expressions,
$F_2^n/F_2^p \approx F_1^n/F_1^p \approx g_1^n/g_1^p \to \mu_n^2/\mu_p^2$
as $Q^2 \to \infty$.
On the other hand, for the $g_2$ structure function, which depends on
both $G_E$ and $G_M$ at large $Q^2$, one has a different asymptotic
behavior, $g_2^n/g_2^p \to \mu_n^2 / (\mu_p (\mu_p-1)) \approx 0.73$.

Of course the reliability of the duality predictions are only as good
as the quality of the empirical data on the electromagnetic form factors.
While the duality relations are expected to be progressively more accurate
with increasing $Q^2$ \cite{RUJ}, the difficulty in measuring form factors
at large $Q^2$ also increases.
Experimentally, the proton magnetic form factor $G_M^p$ is relatively well
constrained to $Q^2 \sim 30$ GeV$^2$, and the proton electric $G_E^p$ to
$Q^2 \sim 10$ GeV$^2$.
The neutron magnetic form factor $G_M^n$ has been measured to
$Q^2 \sim 5$ GeV$^2$, although the neutron $G_E^n$ is not very well
determined at large $Q^2$ (fortunately, however, this plays only a minor
role in the duality relations, with the exception of the neutron to
proton $g_2$ ratio, Eq.(\ref{SFdualg2})).

In Fig.~1 the solid curves represent the $A_1$ asymmetry calculated from
actual form factor data, while the dashed extensions illustrate the
extrapolation of the form factors in Ref.\cite{MDM} beyond the currently
measured regions of $Q^2$.
The fit in Ref.\cite{MDM} uses all the available form factor data at
lower $Q^2$, together with perturbative QCD constraints beyond the
measured region at large $Q^2$.
To test the sensitivity of the results in Figs.~1 and 2 to form factor
shapes we have used several different parameterizations from
Refs.\cite{BOSTED,GK}, as well as a pure dipole form.
Compared with the latter, the proton polarization asymmetries in Fig.~1
vary by $\sim$ 18\%, 7\% and 2\% at $Q^2 = 1, 10$ and
50~GeV$^2$, respectively, corresponding to values of $x$ at the pion
threshold of $x_{\rm th} = 0.78, 0.97$ and 0.99, respectively.
The neutron asymmetries vary by $\sim$ 100\%, 14\% and 3\% at the same
values (the large relative variation at $Q^2=1$~GeV$^2$ simply reflects
the fact that $A_1^n \approx 0$ at $x_{\rm th} \sim 0.8$).
As would be expected, the uncertainties decrease with increasing $Q^2$,
since both the dipole fit and the fit from Ref.\cite{MDM} incorporate
the correct $Q^2 \to \infty$ limits from perturbative QCD.

The differences between the neutron to proton structure function ratios in
Fig.~2 and those calculated from dipole parameterizations of form factors
are qualitatively similar to those for the polarization asymmetry, namely
$\sim 6\%$ and 25\% for $F_2$ and $g_1$, respectively,
at $Q^2 = 5$~GeV$^2$, decreasing to $\sim 4\%$ and 10\%
at 20~GeV$^2$.
We have also tested the sensitivity of the ratios to the new data from
Jefferson Lab on the $G_E^p/G_M^p$ ratio \cite{JLABGE}, which show
deviations from dipole behavior for $Q^2 \alt 3.5$~GeV$^2$.
The differences induced in the ratios in Figs.~1 and 2 are, however,
within the quoted range of uncertainty.

Obviously more data at larger $Q^2$ would allow more accurate predictions
for the $x \to 1$ structure functions, and new experiments at Jefferson
Lab \cite{JLABGE} and elsewhere will provide valuable constraints.
However, the most challenging aspect of testing the validity of the local
duality hypothesis is measuring the inclusive structure functions at high
enough $x$.
Rapidly falling cross sections as $x \to 1$ mean that only very high
luminosity facilities will be able to extract these with sufficiently
small errors.
The most promising possibility at present is the energy-upgraded CEBAF
accelerator at Jefferson Lab.
Once data on the longitudinal and spin-dependent structure functions at
large $x$ become available, a more complete test of local duality between
elastic form factors and $x \sim 1$ structure functions can be made.
In particular, with data on both the $F_1$ and $F_2$ (or $g_1$ and $F_2$)
structure functions at large $x$ one will be able to extract the $G_E$
and $G_M$ form factors separately, without having to assume the $G_E/G_M$
ratio in extracting $G_M$ from the currently available $F_2$
\cite{JLAB,RUJ}.

Along with the spin dependence, unraveling the flavor dependence of
duality is also of fundamental importance.
Although the local duality relations discussed here are empirical,
a more elementary description of the quark--hadron transition requires
understanding the transition from coherent to incoherent dynamics
and the role of higher twists for individual quark flavors.
This is as relevant for all the $N \to N^*$ transition form factors
as for the elastic.
The flavor dependence can be determined by either scattering from
different hadrons, or tagging mesons in the final state of semi-inclusive
scattering in the resonance region.
Unraveling the flavor and spin dependence of duality, and more generally
the relationship between incoherent (single quark) and coherent
(multi-quark) processes, will shed considerable light on the nature of
the quark $\to$ hadron transition in QCD.

\acknowledgements

I would like to thank S.J.~Brodsky, C.E.~Carlson, F.E.~Close, R.~Ent,
N.~Isgur, S.~Jeschonnek, X.~Ji, C.~Keppel, F.M.~Steffens, A.W.~Thomas
and J.W.~Van~Orden for helpful discussions, and K.~Tsushima for drawing
my attention to typographical errors in Eqs.~(2).
This work was supported by the Australian Research Council, and DOE
contract DE-AC05-84ER40150.


\begin{figure}
\epsfig{figure=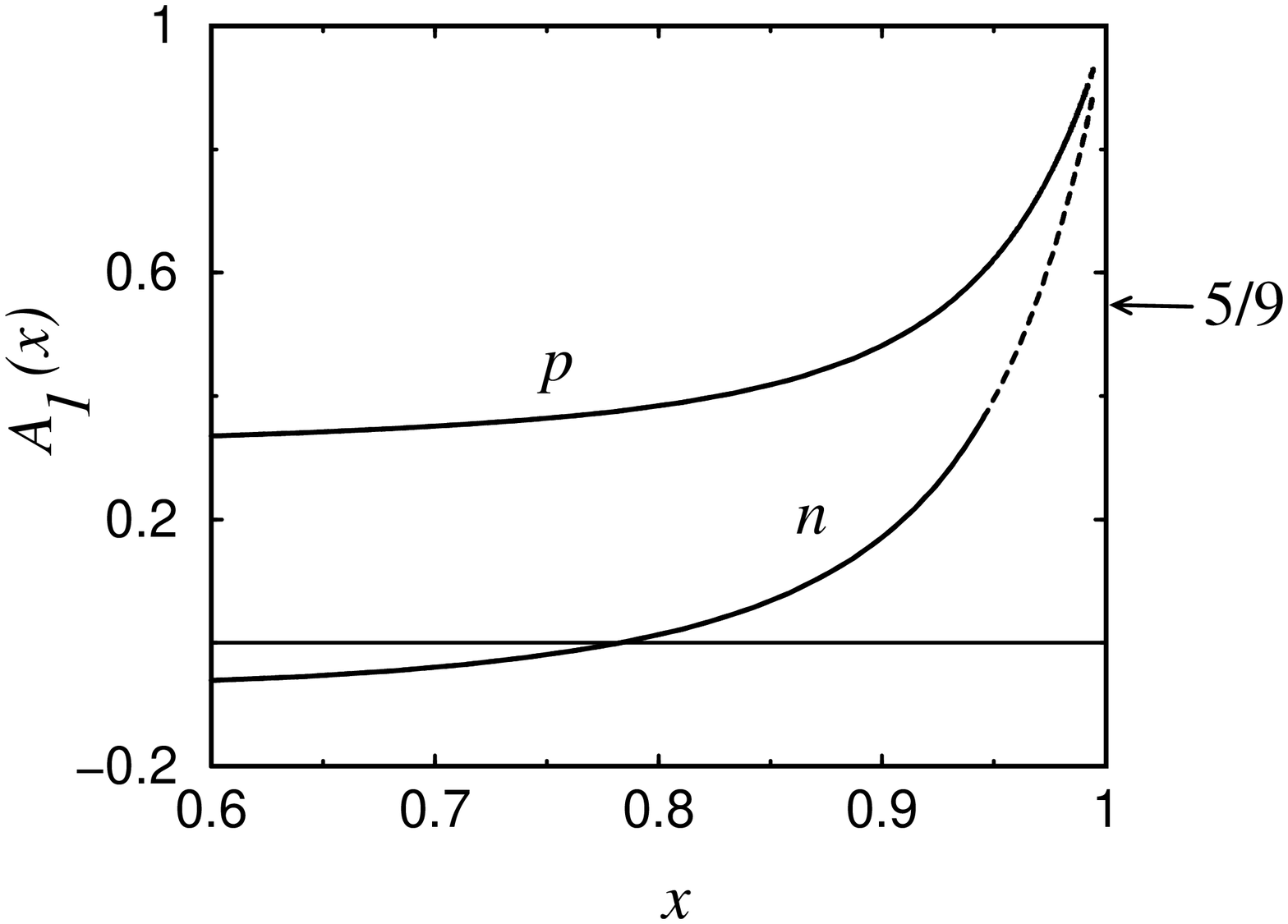,height=11cm}\\
\caption{Polarization asymmetries $A_1$ for the proton and neutron
	at large $x$.
	The SU(6) predictions are 5/9 for $p$ and 0 for $n$.
	The dashed extensions represent asymmetries calculated from
	extrapolations of form factors beyond the currently measured
	regions of $Q^2$.}
\end{figure}

\newpage
\begin{figure}
\epsfig{figure=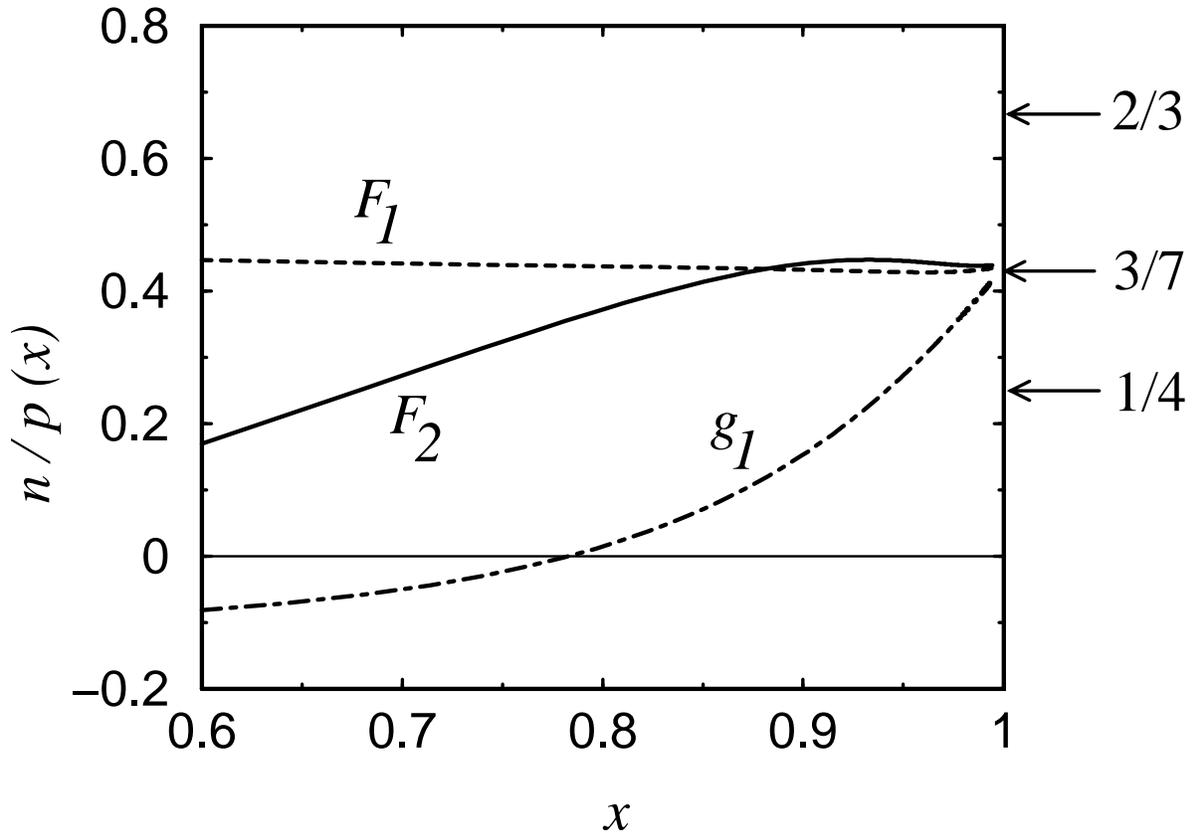,height=11cm}\\
\caption{Neutron to proton ratio for $F_1$ (dashed), $F_2$ (solid)
	and $g_1$ (dot-dashed) structure functions at large $x$.
	Several leading twist model predictions for $F_2$ in the
	$x \to 1$ limit are indicated by the arrows: 2/3 from SU(6),
	3/7 from SU(6) breaking via helicity retention, and 1/4 from
	SU(6) breaking through $d$ quark suppression.}
\end{figure}

\end{document}